\documentstyle[eqsecnum,array,multicol,prl,aps,epsf]{revtex}
\draft
\tighten
\begin{document}

\title{DYNAMICAL CHAOS AND POWER SPECTRA IN TOY MODELS OF 
HETEROPOLYMERS AND PROTEINS}

\author{Mai Suan Li$^{1,2}$, Marek Cieplak$^1$ and Nazar Sushko$^1$}

\address{$^1$Institute of Physics, Polish Academy of Sciences,
Al. Lotnikow 32/46, 02-668 Warsaw, Poland\\
$^2$ Institut f{\H u}r Theoretische Physik, Universit{\H a}t zu K{\H o}ln,
Z{\H u}lpicher Stra{\ss}e 77, D-50937 K{\H o}ln, Germany }

\address{
\centering{
\medskip\em
{}~\\
\begin{minipage}{14cm}
The dynamical chaos in Lennard-Jones toy models of heteropolymers is studied
by  molecular dynamics simulations. It is shown that two nearby trajectories
quickly diverge from each other if the 
heteropolymer corresponds to a random sequence.
For good folders, on the other hand, two nearby trajectories may initially move
apart but eventually they come together. 
Thus good folders are intrinsically non-chaotic.
A choice of a distance of
the initial conformation from the native state affects the way in which
a separation between the twin trajectories behaves in time. This observation
allows one to determine the size of a folding funnel in good folders.
We study the energy landscapes of the toy models by determining
the power spectra and fractal characteristics of the 
dependence of the potential energy on time.
For good folders, folding and unfolding trajectories have distinctly
different correlated behaviors at low frequencies.
{}~\\
{}~\\
{\noindent PACS Nos. 71.28.+d, 71.27.+a}
\end{minipage}
}}

\maketitle

\begin{multicols}{2}

\section{Introduction}

The notion of chaos in physical systems has several meanings. For instance,
in the context of spin glasses it refers to the phenomenon of 
instability of the ground state against 
weak perturbations in the exchange couplings \cite{Bray}.
In the context of dynamical systems, on the other hand, it refers to 
an acute sensitivity of trajectories to the initial conditions.
Both of these meanings have relevance for understanding proteins and random
heteropolymers.

In sequences of aminoacids, the static chaos can be probed by investigating
how mutations affect stability of the ground state. Several
studies \cite{Banavar,Thirumalai,Shakhnovich} have demonstrated that
the native state of a random heteropolymer is unstable against mutations
whereas there is stability in designed sequences provided the
lengths of the sequences, $N$, are sufficiently small.
It should be noted that the mutations may
have strong effect on folding kinetics even in the case of short proteins
\cite{Pande}. Investigation of the effects of an aminoacid substitution
in a protein is an essence  of experimental procedures aimed at determining
the transition state in folding\cite{Fersht}.

Here, we focus on the notion of dynamical chaos in such systems and ask how
do two folding trajectories relate to each other
as the system evolves from two nearby conformations.
Furthermore, can such information provide clues about the nature
of the energy landscape of the system?

We consider continuum space models, as opposed
to lattice models, since the latter have intrinsically
discretized dynamics of a rather arbitrary nature.
Specifically, we consider three toy $N$=16 off-lattice
models that have been extensively characterized before \cite{Li,Li1,Hoang}
and whose native
states are shown in Figure 1. The Hamiltonians of these systems
are defined in terms of the Lennard-Jones potentials
(see Sec.2) and the time evolution is determined by Newton's equations
that are solved by using the methods of molecular dynamics \cite{MD}.
The first two of these systems, denoted in \cite{Li,Li1} as
$G$ and $R'$, are two-dimensional  whereas the third,
denoted by $H$, is a Go-like model \cite{Go} of a three-dimensional
helix \cite{Hoang}. 
The quality of folding of these systems has been determined by using
thermodynamic \cite{Camacho} and kinetic \cite{Socci} criteria.
Among the three systems, $G$ is found to be a good folder, 
$R'$ is a bad folder, and $H$ has itermediate folding properties. 
Thus $G$ is an analog of a 
protein and $R'$ corresponds to a typical random heteropolymer and
$H$ is a borderline case. The task of this paper is to compare
the chaos related properties across this range of foldability.

As a measure of the distance between two conformations $a$ and $b$
we take $\delta_{ab}$ where
\begin{equation}
\delta^2_{ab} \; \; = \; \; \frac{1}{N^2-3N+2} \sum_{i \neq j,j\pm 1}
(| \vec{r}_i^a - \vec{r}_j^a | - | \vec{r}_i^b - \vec{r}_j^b |)^2 \; \;, 
\end{equation}
and where $\vec{r}_i^a$ is the position vector of the $i$'th monomer
in the conformation $a$. This distance involves relative distances 
between the monomers and 
is bounded due the finite spacial extent of any conformation.
Thus this distance cannot diverge as time grows which makes the 
characterization of the dynamical chaos unconventional in this case.

Notice that once the system folds from a random conformation
to the native state, it may either stay in the immediate vicinity  
of the native state or it may depart further away and then keep visiting
the native state. 
This translates into two possible scenarios for the asymptotic behavior
of $\delta _{ab}$:
either the distance saturates asymptotically 
at a finite value -- the case
of $H$ and $R'$ -- or it tends to zero -- the case of $G$.
Thus good folding properties are reflected in the small asymptotic values
of $\delta _{ab}$ which may be used as an alternative criterion for good
foldability. The asymptotic saturation of $\delta _{ab}$ in bad folders
is achieved  at a much shorter time scale than one needed to establish
the asymptotic tendency in good folder. Thus the bad folders can be said
to be more chaotic than the good folders.


Studies of the distance between two trajectories provide information
about the energy landscape available to a sequence. A complementary
information can be obtained by studying individual trajectories. This is
shown in Section 4 where we consider fractal and spectral properties
of the potential energy curve, $E_p(t)$,
as seen on a trajectory as a function of time, $t$.
The power spectrum, obtained by Fourier transforming
$E_p(t)$, is found to 
indicate a correlated noise pattern and it shows sensitivity
to a sequence in a way which is consistent with $H$ being intermediate.
Furthermore, the corresponding low frequency
power law exponent is found to be
markedly different for folding and unfolding trajectories in 
systems of good foldability.
The fractal dimensionality of the $E_p(t)$ curve is determined according
to a procedure developed in \cite{Thirumalai1,Dubuc}.
We find that, for the well folding system, the temperature
dependence of this fractal dimensionality has 
a dip around a temperature which is optimal for folding.
No such dip arises in poor folders.

\section{The sequences}

We start our discussion by defining models that we study.
The sequences denoted by $G$ and $R'$ are
two-dimensional
versions of the model introduced by Iori, Marinari and Parisi \cite{IMP}.
Their native states are shown in Figure 1.
The Hamiltonian is given by
\begin{equation}
H\; \; = \; \; \sum_{i \neq j} \{ k (d_{i,j} - d_0)^2 \delta_{i,j+1}
+ 4 \epsilon [ \frac{C}{d_{i,j}^{12}} - \frac{A_{ij}}{d_{i,j}^6} ] \} \; \; ,
\end{equation}
where $i$ and $j$ range from 1 to 
$N$=16. The distance between the beads,
$d_{i,j}$, is defined as $ |\vec{r}_{i}-\vec{r}_{j}|$, where
$\vec{r}_i$ denotes the position of bead $i$. 
$d_{ij}$ is measured in units of $\sigma$ the typical value of which is
$\sigma=5\r{A}$.
The harmonic term
in the Hamiltonian, with the spring constant $k$,
couples the beads that are adjacent along the chain.  The remaining terms 
represent the Lennard-Jones potential. 
In \cite{IMP}, $A_{ij} = A_0 + \sqrt{\beta}\eta_{ij}$, 
where $A_0$ is constant
and $\eta_{ij}$'s are Gaussian variables with zero mean and unit variance;
$\beta$ controls the strength of the quenched disorder. 
The case of $\eta_{ij}=0$ and $A_0= C$ would correspond to a
homopolymer with the standard Lennard-Jones interaction used in 
simulations of liquids. In Eq. (2.1) $\epsilon$ is the typical Lennard-Jones
energy parameter. We adopt the units in which $C$=1 and
consider $k$ to be equal to 25$\epsilon$.
Smaller values of $k$ may violate the self-avoidance of the chain \cite{Li}.
The coupling constants $A_{ij}$ for system $R$' are listed in ref. \cite{Li}.
These are shifted Gaussian-distributed numbers with the strongest attracting
couplings assigned to the native contacts. For system $G$, $A_{ij}$ is taken
to be 1  or 0 for the native and non-native contacts respectively.
System $R$' has been shown to be structurally overconstrained and hard to fold.

\begin{figure}
\epsfxsize=4in
\centerline{\epsffile{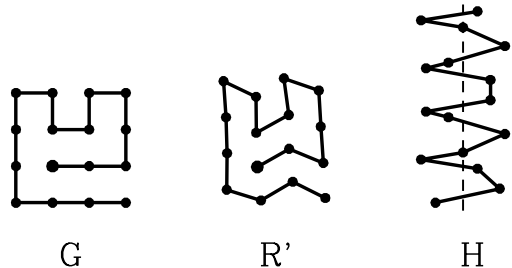}}
\caption{Native conformations of there sequences studied
in this work.}
\end{figure}

The helical system $H$ has a native state, shown in Figure 1, that mimics 
typical $\alpha$-helix secondary structures.
In this case the distances between beeds are assumed to have the
length $d_0=3.8 \r{A}$. 
As one proceeds along the helix axis from one bead to another,
the bead's azimuthal angle is rotated by 100$^o$ and the azimuthal length
is displaced by 1.5 $\r{A}$. The Hamiltonian used to describe the
helix is a Go-like modification of eq.2 and it reads \cite{Hoang}
\begin{equation}
H \; \; = \; \; V^{BB} \; + \; V^{NAT} \; + \; V^{NON} \; .
\end{equation}
The first term is a backbone potential which includes the harmonic and 
anharmonic interactions
\begin{equation}
V^{BB} \; = \; \sum_{i=1}^{N-1} [ k_1 (d_{i,i+1} -d_0)^2 
+  k_2 (d_{i,i+1} -d_0)^4 ] \; .
\end{equation}
We take $d_0=3.8 \r{A}$, $k_1=\epsilon$ and $k_2=100\epsilon$.
The interaction between residues which form native contacts in the
target conformation is chosen to be of the Lennard-Jones form
\begin{equation}
V^{NAT} \; = \; \sum_{i<j}^{NAT} 4 \epsilon [(\frac{\sigma _{ij}}{d_{ij}})^{12}
- (\frac{\sigma _{ij}}{d_{ij}})^{6} ] \, .
\end{equation}
We choose $\sigma _{ij}$ so that each contact in
the native structure is stabilized at the minimum of the potential,
i. e. $\sigma _{ij} = 2^{-1/6} d_{ij}^N$, where $d_{ij}^N$ is the 
length of the corresponding native contact.
Residues that not form the native contacts interact via a repulsive soft core 
potential $V^{NON}$, where
\begin{eqnarray}
V^{NON} \; \; = \; \; \sum_{i<j}^{NON} \; V_{ij}^{NON} \; , \\
V_{ij}^{NON}= \left\{ \begin{array}{r@{\quad \quad}l}
4 \epsilon [(\frac{\sigma _0}{d_{ij}})^{12}-(\frac{\sigma _0}{d_{ij}})^6 ]+
\epsilon &
, d_{ij}<d_{cut}\\
0 & , d_{ij}>d_{cut}.
\end{array} \right. 
\end{eqnarray}
Here $\sigma _0=2^{-1/6} d_{cut}, d_{cut}=5.5 \r{A}$.

The time evolution  of the sequences is determined by the fifth
order predictor-corrector scheme \cite{MD}. The integration step
is chosen to be 0.005$\tau$, where $\tau=m \sigma ^2 / \epsilon$
is the characteristic time unit and $m$ is the mass of a bead.
In order to simulate systems in contact with
a heat bath of temperature $T$, we augment the equations of motion by
the Langevin uncorrelated noise terms as described in \cite{Hoang}:
\begin{equation}
m\ddot{{\bf r}} = -\Gamma \dot{{\bf r}} + F_c + \eta \;,
\end{equation}
where $F_c=-\nabla_r E_p$ and
\begin{equation}
\left<\eta(0)\eta(t)\right> = 2\Gamma k_B T \delta(t),
\label{eqgam}
\end{equation}
where $k_B$ is the Boltzmann constant. 
We take $\Gamma $ equal to 2. In the following, the temperature will be
measured in the reduced units of $\epsilon/k_B$.

\begin{figure}
\epsfxsize=3.2in
\centerline{\epsffile{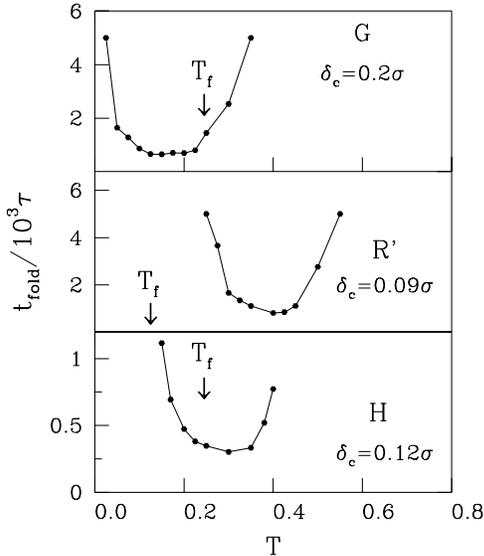}}
\caption{The temperature dependence of the folding times
for sequences $G$, $R'$ and $H$. The
results are based on 100 starting random conformations. The
arrows indicate the folding temperatures.}
\end{figure}
Figure 2 shows the $T$-dependence of folding times for the three
sequences studied.
The folding time is defined as the median first passage time and 
the folding is declared to be accomplished if the distance to the native state, 
$\delta_{ns}$,
becomes smaller than the native basin size $\delta_c$. $\delta_c$ is defined by
the shape distortion method \cite{Li1} and it is equal
to (0.2$ \pm 0.01) \sigma$ and (0.09$ \pm 0.01)\sigma$ 
for sequence $G$ and $R'$, respectively.
For sequence $H$, $\delta _c = (0.6 \pm 0.05) \r{A} = (0.12 \pm 0.01)\sigma$

The folding is the fastest at the temperature $T_{min}$ below which 
glassy kinetics set in.  We determine that
$T_{min}$=0.15 $\pm 0.02$, 0.4 $\pm 0.02$, and 0.3 $\pm 0.02$
for $G$, $R'$ and $H$ respectively.
Similar estimates of $T_{min}$ were obtained with a  Monte Carlo  "dynamics"
\cite{Li} (at a larger CPU cost).

Socci and Onuchic \cite{Socci} have proposed that what determines good
foldability is  whether  the folding temperature, $T_f$, is outside
of the range of temperatures where kinetics become glassy. $T_f$
is defined as a temperature at which the equilibrium probability of
being in the native state is 1/2. Here, we rephrase this criterion
in the following way:
a sequence is a good folder if $T_f$ is greater than $T_{min}$, otherwise
foldability is bad.  We determine
$T_f$ through a Monte Carlo process  and get
values indicated by the arrows in Figure 2. The relative values of
$T_{min}$ and $T_f$ indicate what was announced in Section 1:  $G$ is 
a good folder, $H$ is intermediate and $R'$ is a bad folder.
Characterization based on the specific heat and structural susceptibility
yields a similar conclusion \cite{Li,Li1,Hoang}.

\begin{figure}
\epsfxsize=2.8in
\centerline{\epsffile{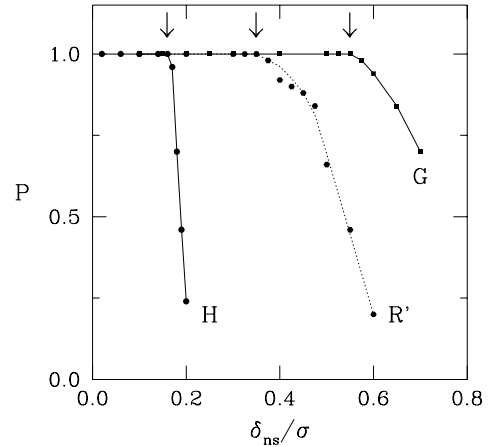}}
\caption{The dependence of the probability, $P$, of falling into the native
state as a result of quenching
on the distance to the native state. The arrows indicate the values of
$\delta _f$. The results are obtained by the Monte Carlo method and
are averaged over 100 starting configurations.}
\end{figure}

Another relevant property of a sequence is its characteristic funnel 
size, $\delta _f$. It can be estimated by generating random conformation
of the system and then quenching them (by setting $T=0$) and determining
whether the resulting quenched state is native or not\cite{Li1}.
This allows one to estimate the probability, $P$, of getting to the
native state from a conformation which is $\delta$ away form the native state.
The critical value of $\delta$, $\delta _c$
above which this probability becomes smaller than one may
be identified as a characteristic $\delta _f$.
Our results on $P$ are shown in Figure 3 from which we  can deduce that
$\delta _f \approx 0.55\sigma$,
0.35$\sigma$ and 0.16$\sigma$ for $G$, $R'$ and $H$, respectively
(for $H$ we took $\sigma=5\r{A}$).
Thus the good folder $G$ has a much larger funnel, as measured by the
distance $\delta _{ns}$, than the bad folder $R'$. A direct comparison
of $\delta _f$ for $G$ and $R'$ to that for $H$ is not meaningful 
because the Hamiltonians and dimensionalities are different.
Thus even though $\delta _f/\sigma$ for $H$ is the smallest  among the
three sequences, its foldability is intermediate.

\section{Dynamical chaos}

In order to study the dynamic chaos we monitor two trajectories which
evolve in time from two conformations which are
initially separated by a
distance $\Delta R$. We chose $\Delta R =0.001 \sigma$. Smaller values 
of $\Delta R$
yield qualitatively similar results.
The forces due to the Langevin noise are identical for both trajectories.

Figure 4 illustrates what happens with two initially nearby
trajectories as a function of time for systems $G$, $R'$, and $H$
respectively. The trajectories are characterized by the
potential energy, radius of gyration, $R_g$, and the number of established
native contacts, $N_c$.  
Two monomers $i$ and $j$ are declared to
form a native contact if the distance between them is in the
interval $[0.9 d_{ij}^N , 1.1 d_{ij}^N$].
Independent of which of these quantities is
used, we see that the two trajectories quickly come together for system $G$
but continue to be clearly distinct for a much longer time in the case of
system $R$'. System $H$ involves larger scales on the $y$-axis and, on a closer
inspection, behaves like $R$'. 
\end{multicols}
\begin{figure}
\epsfxsize=5.6in
\centerline{\epsffile{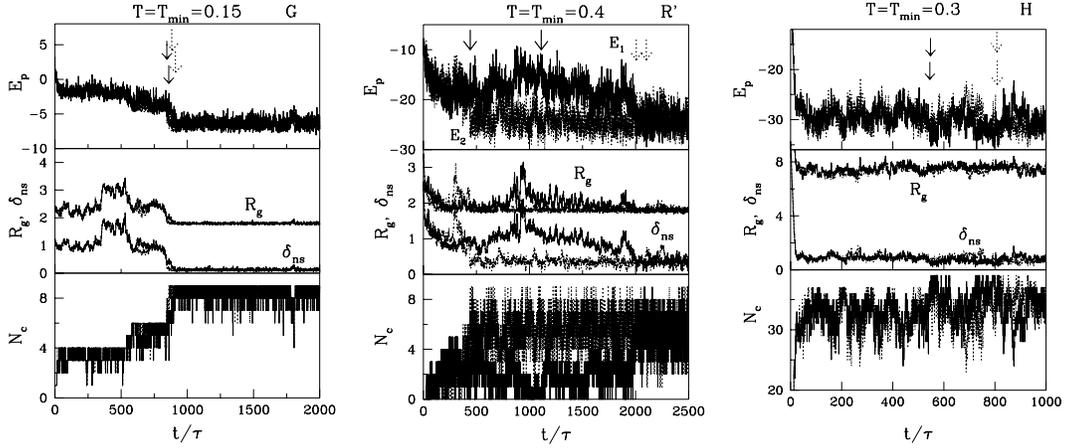}}
\vspace{0.1in}
\caption{Time evolution of the potential energy $E_p$, the gyration radius
$R_g$, the distance to the native state $\delta_{ns}$ and the number
of native contacts $N_c$ for two typical trajectories which  have the initial
departure $\Delta R=0.001$.
The left-hand, center, and right-hand parts of the figure
correspond to sequences
$G$, $R'$, and $H$ respectively  and all data refer to $T=T_{min}$. 
The first and second
trajectory
are denoted by solid and dotted line respectively. The left arrow
indicates when the last native contact appears for the first time, and the
right arrow indicates when folding takes place, i.e.
$\delta_{ns} < \delta_c$.}
\end{figure}

\begin{multicols}{2}
\begin{figure}
\epsfxsize=2.8in
\centerline{\epsffile{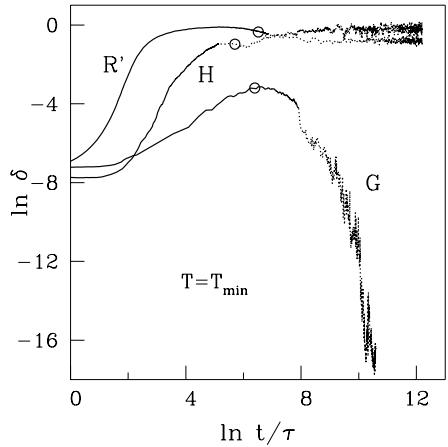}}
\caption{The time dependence of $\delta$ for $G$, $H$ and $R'$
at $T=T_{min}$ for two nearby trajectories which start at two
nearby unfoloded conformations.
The solid portions of the curves are averaged over 1000 to 1500 pairs
of such trajectories. The dotted portions are averaged over 20 to 200
pairs. The circles indicate values of the median folding times. }
\end{figure}

Figure 5 shows the time dependence of $\delta$ at $T=T_{min}$ averaged
over many pairs of starting trajectories. The time scales extend to much
beyond the folding times.
For $R'$ and $H$ the distance $\delta$ remains nonzero at these
large time scales but for $G$ --- $\delta \rightarrow 0$.
Thus for good folders all two initially separated trajectories eventually
come together. This is not so for bad and intermediate folders.
In general, the better the folder, the longer it takes to establish
the asymptotic behavior in $\delta$. We can rephrase it by stating
that $R$' is more chaotic than $H$ because the asymptotic large
separation between the trajectories is established sooner.


Figure 5 refers to trajectories which start in unfolded conformations.
It is interesting to consider what happens when the starting conformations
come from a closer neighborhood of the native state.
We generate such starting "points" by evolving a trajectory from
an unfolded conformation to various stages chracterized by
predefined values of $\delta _{ns}$. At each state, we spawn a
nearby companion trajectory which is
displaced by $\Delta R$ from the current conformation of the leading
trajectory.
The stages can alternatively be characterized by the numbers
of established native contacts but the $\delta _{ns}$ is more
convenient to use when one deals with long heteropolymers.

The first panel in the left-hand part of Figure 6 shows the time dependence 
of $\delta$ for $G$ at $T=0.1$,
0.15, and 0.2 for times significantly shorter than those corresponding
to Figure 5.
It is interesting to point out that the short time behavior of $\delta _{ab}$
may yield information about the size of the folding funnel.
This can be achieved by studying two initially
nearby trajectories which start at various locations in the folding funnel,
i.e. at various distances $\delta _{ns}$ away from the native state,
as illustrated in Figure 6.
Notice that the initial
placement affects the character of the initial evolution of
$\delta$. For $\delta _{ns}\; >\;
\delta _{ns}^{th} =0.55\sigma$ and $T=T_{min}=0.15$ (the first panel 
of Figure 6),
the two trajectories diverge.
Otherwise, they reduce their relative distance. This can be interpreted
as a situation in which the twin trajectories are
placed within the folding funnel. Thus the threshold value of $\delta _{ns}$
under the optimal folding conditions
should be a measure of size of the folding funnel. In fact, this threshold
value agrees with the funnel size as determined from Figure 3.
For temperatures above $T_{min}$ the kinetic conditions deteriorate,
the funnel desintegrates, and the threshold behavior in chaos disappears:
the system becomes more chaotic.
Below $T_{min}$, on the other hand, the funnel also fades away but
the system is close to the quenching conditions. 
As the system evolves it becomes to be
driven by energy minimization so the trajectories start to come
closer together right away even for substantial values of $\delta _{ns}$.

The center and right-hand parts of Figure 6
show results of a similar analysis for systems $R$' and $H$
respectively.
For bad folders, even if the trajectories start approaching each
other at low $T$, as it happens for $H$ (the bottom right-hand panel 
of Figure 6), 
they do not meet asymptotically
because its is unlikely that they will be simultaneously in the same
energy valleys. $H$ displayes some borderline bahavior but there is
no threshold behavior in system $R$' at any temperature: a viable folding 
funnel never forms.
For good folders, however, studying
twin trajectories allows one to establish the geometry of the folding funnel.

\end{multicols}

\begin{figure}
\epsfxsize=5.6in
\centerline{\epsffile{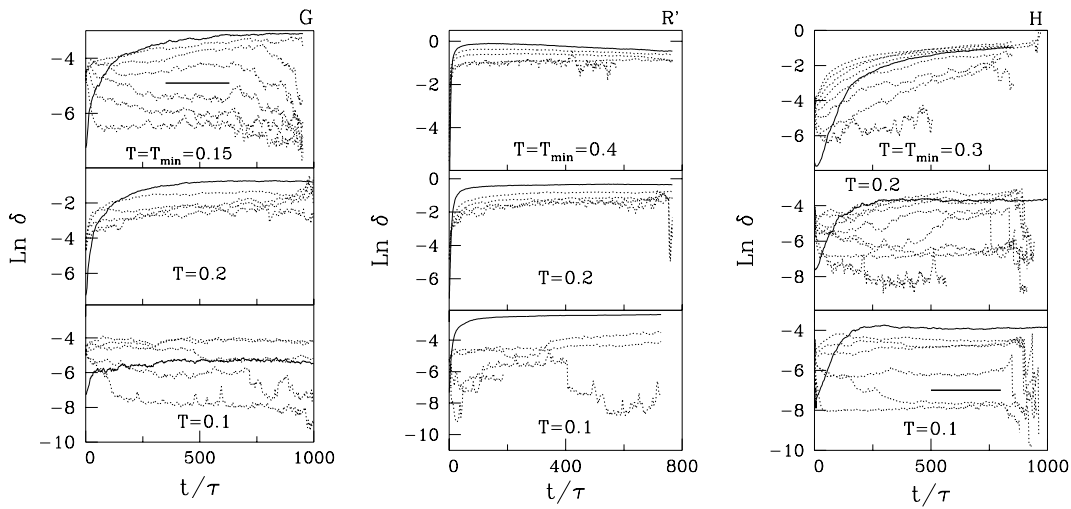}}
\vspace{0.1in}
\caption{The time dependence of $\delta$ for various types
of starting configurations at temperatures which are indicated.
The left-hand, center, and right-hand parts correspond to
sequences $G$, $R$', and $H$ respectively.
For each $T$, the initial separation 
between two trajectories is $\Delta R=0.001\sigma$.
The solid line corresponds to the unfolded
starting configuration.
{\bf Sequence G:}
For $T=T_{min}=0.15$ the dotted lines correspond to
$\delta _{ns}$ = 0.75, 0.6, 0.55, 0.5, 0.35 and 0.2$\sigma$ from top to
bottom respectively.
The curve with $\delta _{ns}^{th}=0.55 \sigma$ has a threshold 
character in the sense  that
for $\delta _{ns}>\delta _{ns}^{th}$ distance $\delta$ decreases with
time monotonically. The bar separates two different types of behavior
of $\delta$.
For $T=0.2$, the dotted lines correspond to
$\delta _{ns}$ = 0.7, 0.5, 0.3, and 0.05$\sigma$ 
from top to bottom respectively.
In the case of $T=0.1$
the dotted lines are arranged in order
$\delta _{ns}$ = 1.6, 1.3, 1.1, 0.8 and 0.5$\sigma$.
For $T=0.2$ the distance between two trajectories grows for any
$\delta _{ns}$. 
{\bf Sequence R':}
For $T=T_{min}=0.4$ the dotted lines correspond
to $\delta _{ns}$ = 0.8, 0.6, 0.4 and 0.05$\sigma$ from  top to bottom
respectively. For $T=0.2$, the dotted lines are arranged in the order
$\delta _{ns}$ = 0.8, 0.6, 0.4, and 0.2$\sigma$.
For T=0.1 the order is
$\delta _{ns}$ = 0.8, 0.7, 0.6 and 0.4$\sigma$.
{\bf Sequence H:}
For $T=T_{min}=0.3$ the dotted
lines correspond to
$\delta _{ns}$ = 2, 1.5, 1.2, 1, 0.8, 0.6, 0.4, 0.3 and 0.15$\sigma$
from top to bottom respectively.
For $T=0.2$
the dotted lines correspond to
$\delta _{ns}$ = 1.2, 1., 0.8, 0.6, 0.5, 0.4, 0.2 and 0.1$\sigma$.
For $T=0.1$
the dotted lines correspond to
$\delta _{ns}$ = 1.5, 1.2, 1.0, 0.8, 0.6, 0.5, 0.4, and 0.2$\sigma$.
The results are averaged over 500 - 3000 starting configurations.}
\end{figure}

\begin{multicols}{2}

In the case of $R'$, for $T=0.2$,
the time dependence of $\delta$ is qualitatively
the same as for $T=T_{min}=0.4$.
The qualitative change is observed, however, at $T=0.1$.
Namely, the apparent
decrease of the distance is seen for $\delta _{ns}^{th}$=0.4 $\sigma$.
It should be noted that the difference
between good folder $G$ and bad folder $R'$ 
is clearly seen only at $T_{min}$. Away from $T_{min}$ they may
behave qualitatively in the same way.

\section{Time evolution of the potential energy}

In order to understand the difference in the chaotic behavior of good
and bad folders better we consider the time dependence
of the potential energy, $E_p$, on individual trajectories under the
optimal folding conditions, i.e. at $T_{min}$.
We consider folding and unfolding trajectories separately and demonstrate
that the corresponding properties of $E_p$ are distinct.

\begin{figure}
\epsfxsize=3.2in
\centerline{\epsffile{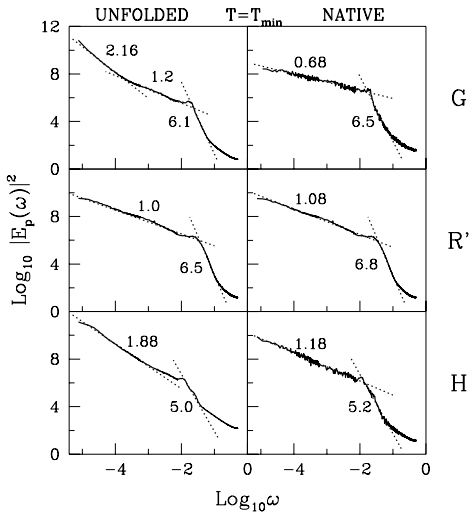}}
\caption{Spectrum of the potential energy for $G$, $R'$ and $H$
at $T_{min}$. The starting conformations are either unfolded
or native as indicated at the top of the figure.
The results are averaged over 400 trajectories.
The values of $f$ are shown next to the curves.}
\end{figure}

The first question we ask is what are the properties of the power spectra
of $E_p$. Figure 7 shows the frequency dependence of $|E_p (\omega)|^2$, where
\begin{equation}
E_p(\omega) \; \; = \; \; \frac{1}{2\pi} \int \exp(i\omega t) E_p(t) dt \; .
\end{equation}
The first observation is that if $E_p$ is viewed as noise then this noise
is clearly correlated -- there is no frequency regime which would
correspond to white noise. In other words, if
\begin{equation}
|E_p|^2 \; \sim \; \omega ^{-s}
\end{equation}
then the exponent $s$ is non-zero.
For each of the systems studied we observe two regimes with a power law
dependence on $\omega$: the low and high frequency regimes which are
separated by $\omega _0 \;\sim\; 0.032 /\tau$ which corresponds
to a time scale of about 200$\tau$.
The values of the power law exponent are indicated in the Figure.
For system $G$, there is also an
intermediate time scale in the folding trajectories where $s=1.20 \pm 0.20$.
It is possible that the existence of the intermediate frequency regime
is a signature of a good foldability in general.

The high frequency behavior corresponds to a large exponent in the power 
law, of order 5 -- 7 (the error bars here are of order 0.5).
The low frequency, i.e. long time
behavior, however, 
is clearly distinct for $G$ and almost the same for $R$'.
For $G$, the low frequency behavior of $E_p$ corresponds to $f$=$2.16 \pm 0.20$
and $0.68 \pm 0.2$ for the folding and unfolding trajectories respectively.
For $R$', on the other hand, one gets the $1/f$ noise for both kinds
of trajectories. This clearly indicates lack of any folding direction in
$R'$.
System $H$ has an intermediate behavior again:
for the low  frequency folding trajectories $s=1.88 \pm 0.24$
and for the unfolding trajectories $s=1.18 \pm 0.20$.
Thus there is a difference between folding and unfolding but the
difference is not as strong as for $G$.

\begin{figure}
\epsfxsize=3.2in
\centerline{\epsffile{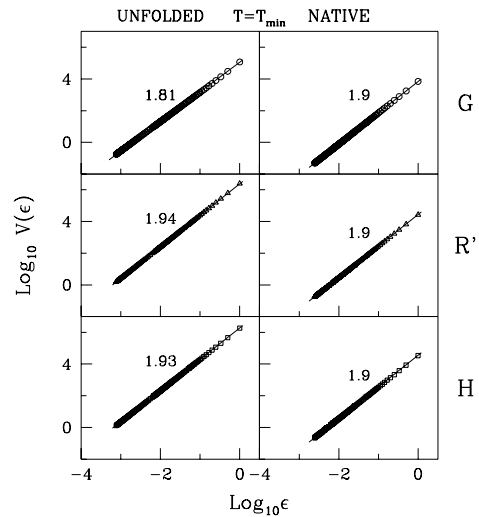}}
\caption{Dependence of $V(\epsilon)$ on $\epsilon$ for $G$, $R'$ and $H$.
The results are obtained at $T=T_{min}$. The left and right hand panels are for
folding and unfolding conformations respectively.
The time steps are equal to the folding times at $T=T_{min}$.
The values of $\gamma$ are shown next to the data points.}
\end{figure}

Another way to characterize trajectories has been recently proposed by
Lidar et al. \cite{Thirumalai1} and it
involves determination of the fractal dimensionality, $\gamma $,
which relates to the self-affinity properties of the $E_p(t)$ curve.
This fractal dimensionality may be obtained by the $\epsilon$-variation
method developed by Dubuc  {\em et al.} \cite{Dubuc}.
Typically, for any function $g$ one can introduce its $\epsilon$-variation
$V(\epsilon ,g)$ as follows
\begin{equation}
V(\epsilon ,g) \; \; = \; \; \int_{0}^{1} v(x,\epsilon ) dx \; ,
\end{equation}
where the $\epsilon$-variation is
\begin{eqnarray}
v(x,\epsilon ) \; = \; \sup _{x'\in R_{\epsilon}(x)}g(x') -
 \inf _{x'\in R_{\epsilon} (x)} g(x') \; , \nonumber\\
R_{\epsilon}(x)=\{ s \in [0,1]; |x-s| < \epsilon \} \; .
\end{eqnarray}

Figure 8 shows the $\epsilon$-dependence of $\epsilon$-variation $V(\epsilon)$ 
for $E_p(t)$ of three typical trajectories at $T_{min}$ run in the
folding or unfolding modes. 
In the folding mode, the trajectories are evolved until the folding is
accomplished. In the unfolding mode,
the trajectories are analysed for a duration of a typical median folding
time at $T_{min}$.
The slopes in Figure 8 give the values of $\gamma$. 
If the starting conformation is native then $\gamma$ appears to be system 
independent 
and equal to about 1.90. 
This again indicates a correlated behavior since for Gaussian distributed
numbers $\gamma$=2.
On the other hand, if the starting conformation
is unfolded, then
$\gamma \approx 1.81$,
1.94 and 1.93 for $G$, $R'$ and $H$ respectively. This suggests that
the smaller the $\gamma$ the better the foldability (and less chaos).

\begin{figure}
\epsfxsize=3.2in
\centerline{\epsffile{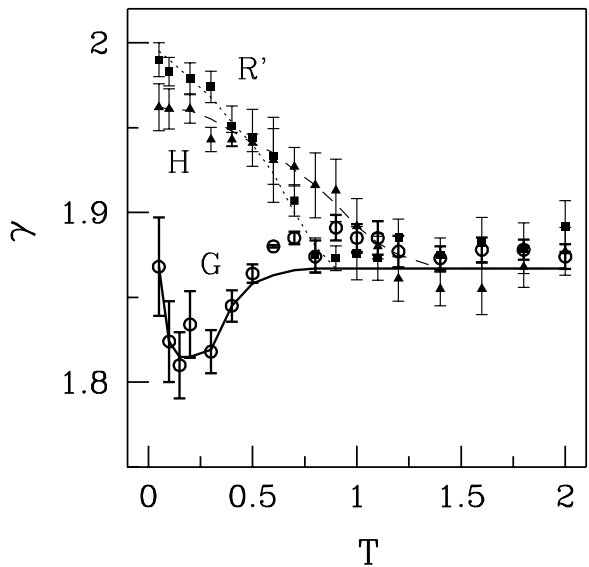}}
\caption{Temperature dependence of $\gamma$ for $G$
(open circles), $R'$ (closed squares) and $H$ (closed triangles).
The starting conformations are unfolded. The time steps are equal
to the folding times at $T=T_{min}$. The number of starting
configurations is 20 - 100.}
\end{figure}

From Figures 7 and 8 one can see that $s$ has a stronger 
system-dependence compared to $\gamma$.
This is due to the fact that the spectral and
roughness properties refer to
different aspects of the behavior.
This may be seen clearly in the case of the white noise where
$\gamma =2$ but $s=0$ which means that the white noise is not correlated
but its profile remains rough. In other words the spectral analysis
provides information about the pattern correlation whereas
the fractal dimensionality relates to the roughness.
Therefore, $\gamma$ and $s$ may depend on the system in a different way.
It is interesting to ask why are the system-dependences of  $s$ and $\gamma$
in the folding mode 
stronger compared to the unfolding mode.
The reason seems to be that the time
dependence of the potential energy in the folding mode
(shown in Figure 4) is substantially
stronger than in the unfolding mode
for all of the three sequences. Thus, in the unfolding
case the system-dependence becomes weaker and $\gamma$ even loses
the dependence entirely. It would be interesting
to know if this observation is still valid in real proteins.

The temperature dependence of $\gamma$ for three sequences is shown
in Figure 9 where the starting conformations are unfolded. 
The values of $\gamma$ presented in this figure fulfil the
relation $\gamma = 2 - \alpha$, where $\alpha$ is the roughness exponent
\cite{Thirumalai1}, if $\alpha$ is determined directly.
Studies of models of real proteins
\cite{Thirumalai1} indicate that $\gamma$ may depend on $T$ weakly.
This is also true for our model systems as shown in Figure 9.
Note that system $H$ again behaves in a way which is intermediate
between $R$' and $G$.
Interestingly, in the protein-like sequence $G$ we observe a dip in
$\gamma$ around $T_{min}$. A similar but wider
dip was also observed for real proteins
like myoglobin, BPTI and PPT \cite{Thirumalai1}.
The presence of the dip could be explained in the following way:
around $T_{min}$ the system establishes a folding funnel
and the motion becomes less rugged or less chaotic.
Thus, fractal analysis around $T_{min}$ may provide 
the useful information about the foldability. It should be noted
that our results have been obtained for times equal to
the folding times at $T_{min}$ and the related conclusions are valid
on these time scales. Longer or
shorter runs may, in principle,
change the estimates of $\gamma$ \cite{Thirumalai1}.
Within the error bars, the results shown in Figure 9, however, 
do not change if the the time scale is doubled.


In conclusion, we have studied the dynamic chaos of several model 
sequences and demonstrated that good folders are essentially
non-chaotic and bad folders are intrinsically chaotic.
The energy landscape of heteropolymers can be characterized by
the spectral and fractal properties of the time evolution of the
potential energy of the system.

We thank T. X. Hoang for useful discussions and technical help.
This work was supported by Komitet Badan Naukowych (Poland; Grant number\
2P03B-146 18).

\vspace{1cm}


\end{multicols}
\end{document}